\documentclass[conference]{IEEEtran}
\IEEEoverridecommandlockouts
\usepackage{cite}
\usepackage{amsmath,amssymb,amsfonts}
\usepackage{algorithm, algorithmic}
\usepackage{caption}
\usepackage{subcaption}
\usepackage{makecell} 
\usepackage{graphicx}
\usepackage{textcomp}
\usepackage{xcolor}
\usepackage[hidelinks,colorlinks=true,linkcolor=blue,citecolor=blue]{hyperref}
\usepackage{tikz}
\usetikzlibrary{arrows.meta, shapes, shapes.geometric,arrows,fit,matrix,positioning,chains}
\usepackage{pgfplots}
\usepackage{smartdiagram}
\usesmartdiagramlibrary{additions}
\usepackage[draft]{minted}
\usepackage{url}
\setminted{fontsize=\scriptsize}
\makeatletter
\AtBeginEnvironment{minted}{\dontdofcolorbox}
\def\dontdofcolorbox{\renewcommand\fcolorbox[4][]{##4}}
\makeatother
\definecolor{green_keyword}{HTML}{007f00}
\def\BibTeX{{\rm B\kern-.05em{\sc i\kern-.025em b}\kern-.08em
    T\kern-.1667em\lower.7ex\hbox{E}\kern-.125emX}}

\tikzset{
box/.style args = {#1/#2/#3}{rectangle,
        minimum width=#1, fill=#2!30, draw,
        text width =\pgfkeysvalueof{/pgf/minimum width}-2*\pgfkeysvalueof{/pgf/inner xsep},
        minimum height=#3, align=center,
        font=\footnotesize},
box/.default = 18mm/green/0.5cm,
}

\begin{document}
\title{Enabling Retargetable Optimizing Compilers for Quantum Accelerators via a Multi-Level Intermediate Representation\\
\thanks{This manuscript has been authored by UT-Battelle, LLC under Contract No. DE-AC05-00OR22725 with the U.S. Department of Energy. The United States Government retains and the publisher, by accepting the article for publication, acknowledges that the United States Government retains a non-exclusive, paid-up, irrevocable, world-wide license to publish or reproduce the published form of this manuscript, or allow others to do so, for United States Government purposes. The Department of Energy will provide public access to these results of federally sponsored research in accordance with the DOE Public Access Plan. (http://energy.gov/downloads/doe-public-access-plan).}
}

\author{\IEEEauthorblockN{Thien Nguyen}
\IEEEauthorblockA{\textit{Quantum Science Center} \\
\textit{Oak Ridge National Laboratory}\\
Oak Ridge, TN, United States \\
nguyentm@ornl.gov}
\and
\IEEEauthorblockN{Alexander McCaskey}
\IEEEauthorblockA{\textit{Quantum Science Center} \\
\textit{Oak Ridge National Laboratory}\\
Oak Ridge, TN, United States \\
mccaskeyaj@ornl.gov}
}

\maketitle
\thispagestyle{plain}
\pagestyle{plain}

\begin{abstract}
We present a multi-level quantum-classical intermediate representation (IR) that enables an optimizing, retargetable, ahead-of-time compiler for available quantum programming languages. To demonstrate our architecture, we leverage our proposed IR to enable a compiler for version 3 of the OpenQASM quantum language specification. We support the entire gate-based OpenQASM 3 language and provide custom extensions for common quantum programming patterns and improved syntax. Our work builds upon the Multi-level Intermediate Representation (MLIR) framework and leverages its unique progressive lowering capabilities to map quantum language expressions to the LLVM machine-level IR. We provide both quantum and classical optimizations via the MLIR pattern rewriting sub-system and standard LLVM optimization passes, and demonstrate the programmability, compilation, and execution of our approach via standard benchmarks and test cases. In comparison to other standalone language and compiler efforts available today, our work results in compile times that are 1000x faster than standard Pythonic approaches, and 5-10x faster than comparative standalone quantum language compilers. Our compiler provides quantum resource optimizations via standard programming patterns that result in a 10x reduction in entangling operations, a common source of program noise. Ultimately, we see this work as a vehicle for rapid quantum compiler prototyping enabling language integration, optimizations, and interoperability with classical compilation approaches. 
\end{abstract}

\begin{IEEEkeywords}
quantum computing, quantum programming, quantum simulation, programming languages
\end{IEEEkeywords}

\section{Introduction}
Quantum acceleration of existing scientific computing workflows has the potential to enhance computational scalability for modeling and simulation tasks in fields such as nuclear physics, chemistry, and machine learning \cite{Dumitrescu2018,McCaskey2019,hamilton}. As hardware architectures continue to scale and improve --- enabling more qubits at lower error rates --- one can expect quantum-classical machine models to move toward tighter integration of CPU and QPU resources \cite{McCaskeyICRC2018,britt2017high,bertels}. These architectures stand to benefit from robust language and compilation approaches that enable high-level classical language integration, quantum and classical compiler optimization techniques \cite{nam_automated_2018}, and compiler-automated circuit synthesis strategies \cite{younis2020qfast,smith2021leap}. There is a critical need to move the quantum programming community from manual circuit construction via vendor-provided data structures and frameworks embedded in application-level classical languages (Python, etc.) toward performant language approaches that enable tight integration with existing classical runtimes, libraries, and languages. Recently, a number of such language approaches have begun to bridge this gap in the research community, with stand-alone languages such as Q\# \cite{qsharp}, OpenQASM 3.0 \cite{openqasm3, cross2021openqasm}, Silq \cite{silq}, Scaffold \cite{scaffold}, and classical language extensions like QCOR \cite{mintz2019qcor, qcor-acm}. 

In parallel to quantum programming research and development, there has been a wealth of work done to improve classical compilation frameworks and techniques. One result of note is the introduction of \textbf{\emph{multi-level intermediate representations}} enabling compiler representations at a variety of abstraction levels --- including those close to the source language itself --- in tandem with associated progressive lowering workflows that take high-level representations to low-level executable object code via a hierarchy of intermediate representation (IR) abstraction. This enables robust compiler development for domain specific languages that retain automated language-level optimizations, transformations, and lowering to machine-level IRs like the LLVM. Treating quantum program expressions as stand-alone domain specific languages represents an opportunity to leverage these state-of-the-art classical multi-level IRs. Specifically, the MLIR framework \cite{mlir,mlir-doc} is an example of a popular multi-level IR in use today for classically accelerated heterogeneous workflows \cite{mlir-gpu, mlir-nn}, and is well-positioned to provide a unique resource for the rapid prototyping of quantum language compilers via its extensible language-level IR and progressive IR-lowering capabilities. 

We recently demonstrated the utility of MLIR for simple quantum assembly languages with no true control flow structures \cite{mccaskey2021mlir} --- a low level MLIR dialect for quantum computing. In this work, we leverage and extend that simple quantum MLIR extension for a more complex quantum language --- OpenQASM version 3.0 (henceforth referred to as OpenQASM, unless stated otherwise), which provides robust control flow structures, variable declaration and assignment, and novel syntax for quantum circuit generation and synthesis \cite{quantum_modifiers}. Our approach enables an optimizing compiler for OpenQASM that compiles to the LLVM machine-level IR adherent to the recently introduced Quantum Intermediate Representation (QIR) specification \cite{qir,qirspec}. Moreover, this approach need not be limited to OpenQASM --- the implementation pattern shown in this work can serve as a robust mechanism for further quantum language compiler prototyping and deployment. The work presented here puts forward the requisite infrastructure for quantum language expression and LLVM IR generation, leaving future language compiler implementations as a matter of providing the mapping of a language abstract syntax tree (AST) to our quantum MLIR extension (via ANTLR \cite{ANTLR}, for instance). Language lowering to executable code is then readily available.

\begin{figure}[t!]
\begin{tikzpicture}[node distance = 2mm and 2mm, start chain = going right]

\begin{scope}[every node/.style={box, on chain}]
  \node (qDialect)  {Quantum};
  \node (AffineDialect)  {Affine};
  \node (SCFDialect)  {SCF};
  \node (StdDialect)  {Std};
\end{scope}
\node at ($(qDialect.north west)+(0.5,0.15)$) {\tiny{\emph{MLIR Target}}};
\draw[dotted, fill=yellow, fill opacity=0.2] ($(qDialect.north west)+(-0.1,0.35)$) rectangle ($(StdDialect.south east)+(0.1,-0.1)$);
\node (llvmDialect) [box=78.5mm/blue/0.5cm,
                 below=of $(qDialect.south)!0.5!(StdDialect.south)$]
                 {LLVM Dialect - LLVM IR};

\node at ($(llvmDialect.south west)+(1.1,0.25)$) (QIR) [box=20mm/purple/0.35cm, dashed, rounded corners=1mm]{\tiny\textbf{{\emph{QIR Specification}}}};

\begin{scope}[every node/.style={box=25mm/red/0.5cm, on chain}]
  \node at ($(qDialect.north west)+(-0.25,1.5)$) (QASM2) {OpenQASM 2};
  \node (QASM3) {OpenQASM 3};
  \node (Others)  {...};
\end{scope}
\node (openQasm2Parser) [box=25mm/cyan/0.5cm,
                 below=of QASM2]
                 {\tiny{OpenQASM 2 Parser}};
\node (openQasm3Parser) [box=25mm/cyan/0.5cm,
                 below=of QASM3]
                 {\tiny{OpenQASM 3 Parser}};
                 
\node (OtherParser) [box=25mm/cyan/0.5cm,
                 below=of Others]
                 {...};

\node (exe) [box=78.5mm/gray/1.25cm,
                 below=of llvmDialect] {};

\node at ($(qDialect.south west)+(3.85,-1.1)$) {\tiny{\textbf{Binary Executable}}};

\node at ($(qDialect.south west)+(1.5,-1.65)$)  [box=23mm/gray/0.75cm, dashed, rounded corners=0.5mm, align=left, font=\tiny\linespread{0.9}\selectfont]{\tiny\textbf{{\emph{QCOR Runtime Linkage:}}} \\ NISQ / FTQC \newline \newline};

\node at ($(qDialect.south east)+(4.5,-1.65)$)  [box=25mm/gray/0.95cm, dashed, rounded corners=0.5mm, align=left, font=\tiny\linespread{0.9}\selectfont]{\tiny\textbf{{\emph{Classical Compute Codegen:}}} \\ x86, ARM, PowerPC, etc.\newline  \newline    };

\node at ($(qDialect.south west)+(1.45,-1.9)$)  [box=12mm/red/0.25cm, font=\tiny, rounded corners=0.5mm]{\textbf{Accelerator}};

\node at ($(qDialect.south west)+(6.5,-1.9)$)  [box=12mm/red/0.25cm, font=\tiny, rounded corners=0.5mm]{\textbf{Host}};

\draw [thick, black, <->] ($(qDialect.south west)+(2.7,-1.5)$) -- ($(qDialect.south west)+(5.05,-1.5)$);

\end{tikzpicture}
\caption{QCOR's MLIR-based compilation stack for CPU-QPU heterogeneous computing. Each quantum programming language is processed by a dedicated parser that produces the AST of the input source code. AST of different source languages is all mapped to an MLIR representation expressed in the QCOR quantum dialect and built-in Standard, Affine, and SCF dialects.  This MLIR representation is progressively (multi-stage) transformed (optimization and dialect conversion/lowering) until only operations in the LLVM dialect remain, i.e., quantum operations are converted to LLVM function calls adhering to the QIR specification. This guarantees that the final binary executable is compatible with any QIR-conformed runtime implementations provided at link time, such as \texttt{qcor} runtime supporting both remotely hosted (NISQ) and tightly coupled (FTQC) execution models.}
\label{fig:mlir_compile_stack}
\end{figure}
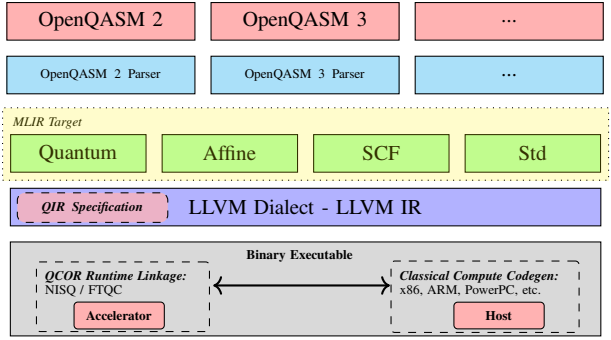
We integrate our approach with the \texttt{qcor} compiler platform \cite{qcor-acm, qcor-github} (note we use QCOR to denote the language extension specification \cite{qcor} and \texttt{qcor} for the compiler implementation). \texttt{qcor} enables single-source quantum-classical programming in both C\texttt{++} and Python, promoting an ahead-of-time C\texttt{++} compiler executable and just-in-time compilation infrastructure for performant quantum-classical code generation and execution. This work extends the \texttt{qcor} executable to stand-alone OpenQASM source files, and enables their compilation and execution in a retargetable (quantum hardware-agnostic) fashion. 

\begin{figure}[t]
\begin{minted}[frame=single,framesep=2pt, fontsize=\footnotesize, breaklines, escapeinside=||, linenos, numbersep=1mm]{llvm}
%Array = type opaque
%Qubit = type opaque

declare void @__quantum__qis__cnot(%Qubit* %0, %Qubit* %1) 
declare void @__quantum__qis__h(%Qubit* %0) 

declare void @__quantum__rt__qubit_release_array(%Array* %0) 
declare i8* @__quantum__rt__array_get_element_ptr_1d( %Array* %0, i64 %1) 
declare %Array* @__quantum__rt__qubit_allocate_array(i64 %0) 

define i32 @ghz() {
  %4 = call %Array* @__quantum__rt__qubit_allocate_array(i64 3)
  %5 = call i8* @__quantum__rt__array_get_element_ptr_1d( %Array* %4, i64 0)
  %6 = bitcast i8* %5 to %Qubit**
  %7 = load %Qubit*, %Qubit** %6, align 8
  call void @__quantum__qis__h(%Qubit* %7)
  %8 = call i8* @__quantum__rt__array_get_element_ptr_1d( %Array* %4, i64 1)
  %9 = bitcast i8* %8 to %Qubit**
  %10 = load %Qubit*, %Qubit** %9, align 8
  call void @__quantum__qis__cnot(%Qubit* %7, %Qubit* %10)
  %11 = call i8* @__quantum__rt__array_get_element_ptr_1d( %Array* %4, i64 2)
  %12 = bitcast i8* %11 to %Qubit**
  %13 = load %Qubit*, %Qubit** %12, align 8
  call void @__quantum__qis__cnot(%Qubit* %10, %Qubit* %13)
  call void @__quantum__rt__qubit_release_array(%Array* %4)
  call void @__quantum__rt__finalize()
  ret i32 0
}
\end{minted}
\caption{A quantum program generating the Greenberger-Horne-Zeilinger (GHZ) state on 3 qubits represented in the LLVM IR adherent to the QIR specification.}
\label{fig:intro_qir}
\end{figure}
\section{Background}
Figure \ref{fig:mlir_compile_stack} demonstrates the hierarchy of layers underlying our compiler architecture. Ultimately, we take quantum-classical languages down to an MLIR representation before emitting standard object code via lowering to the LLVM IR (we leverage the LLVM ecosystem of executables to map an IR bitcode representation to assembly and executable code). Here we seek to provide necessary background on the QIR specification, the MLIR framework, and the \texttt{qcor} compiler frontend and quantum runtime library to set up the presentation of the rest of the compiler architecture.

\subsection{Quantum Intermediate Representation}
Recent work has resulted in the development of a formal specification for a Quantum Intermediate Representation (QIR) embedded in the LLVM IR \cite{qirspec}. This specification does not extend the LLVM IR with new instructions pertinent to quantum computing, instead, it expresses quantum specific operations as declared function calls on opaque data types. Function declarations and corresponding signatures are defined for quantum memory allocation and deallocation, individual qubit addressing, quantum instruction invocation on allocated qubits, and utility functions for array management, tuple creation, and callable invocation. By describing qubits, measurement results, and arrays as opaque types, and promoting function declarations over concrete implementations, the QIR specification promotes a flexible approach to quantum-classical compiler architecture and integration. The approach represents a novel target for language compilers --- all of the LLVM toolchain becomes available, including runtime linking, compile time classical optimizations, and external language interoperability. Figure \ref{fig:intro_qir} demonstrates a LLVM IR code snippet adherent to the QIR specification for the generation of a Greenberger-Horne-Zeilinger (GHZ) maximally entangle state. Note the declaration of opaque \texttt{Array} and \texttt{Qubit} types (lines 1,2) and the externally declared QIR runtime functions (lines 4-10). These are left for implementation by appropriate QIR runtime libraries, affecting actual execution of quantum instructions, array handling (including arrays of \texttt{Qubit} instances), and data type actualization. The body of the code consists of standard LLVM instructions (\texttt{bitcast}, \texttt{call}, etc.) and calls to the declared QIR runtime functions. These functions are concretely provided at link time via the runtime library.

\subsection{MLIR}
Moving up the IR abstraction hierarchy in Figure \ref{fig:mlir_compile_stack}, recent developments in classical compilation research and development has resulted in the MLIR \cite{mlir}. The MLIR represents a modular and extensible approach to defining custom compiler IRs that can express a spectrum of language abstraction (language-level IR down to the machine-level LLVM IR). At its core, the framework puts forward the concept of a language \emph{dialect} which is composed of language-specific \emph{operations}. These operations are the core abstractional unit in the MLIR, and they model a unique mapping of operands to return values, and can optionally carry a dictionary of compile-time metadata (attributes). Operands and return values are modeled as an \texttt{mlir::Value} type, which describes a computable, typed value and its corresponding set of users (enabling one to construct standard use-define chains). The creation of dialects provides a mechanism for mapping language ASTs to a corresponding MLIR operation tree composed of language-dialect-specific operations alongside other utility dialect operations (standard function calls, memory references and allocation, for loops, conditional statements, etc.). Moreover, the framework puts forward a general progressive lowering capability --- incremental translation of higher dialect operations to lower level dialect operations. This enables one to define custom translations from operations in language-level dialects to operations in a machine-level dialects like the LLVM IR. This infrastructure and corresponding workflow provide a flexible architecture for the development of compilation pathways taking language-level syntax trees down to a machine-level IR, like the LLVM.

\subsection{\texttt{qcor}}
\texttt{qcor} provides C\texttt{++}~\cite{qcor-acm} and Python~\cite{qcor-python} language extensions for heterogeneous quantum-classical computing in an effort to promote native quantum kernel programming in a single-source context. Critically, \texttt{qcor} puts forward a compiler runtime library that enables quantum program execution in a multi-modal, retargetable fashion. The execution model enables two mode types of quantum instruction invocation --- \texttt{nisq} and \texttt{ftqc} modes (see QCOR Runtime Linkage in Figure \ref{fig:mlir_compile_stack}). \texttt{nisq} mode supports runtime-level quantum instruction queueing and flushing upon exit of a quantum kernel function, implying a full quantum circuit submission on a remotely hosted quantum computer. \texttt{ftqc} mode models a tightly-coupled CPU-QPU integration model, and quantum instructions are instead streamed as they are invoked, enabling features like fast-feeback on qubit measurement results. The \texttt{qcor} runtime ultimately delegates to the XACC framework \cite{mccaskey2020xacc}, and supports remote execution on IBM, Rigetti, IonQ, and Honeywell quantum processors, and has support for various simulators for the \texttt{ftqc} mode of execution. It is this runtime that plays a critical role in this work, as it provides a target for our QIR runtime library implementation. Moreover, we have provided an entrypoint to OpenQASM compilation natively as part of the \texttt{qcor} command line executable.


\section{Extending OpenQASM 3}
\label{sec:extension}
OpenQASM version 3 has recently been put forward as a formal specification \cite{openqasm3}, and a extended Bachus-Naur form description of the language has been made public as an ANTLR grammar file. This language departs from the previous version (version 2.0) in the introduction of classical control flow and variable declarations, making version 3 much more friendly to hybrid quantum-classical programming. The language provides standard quantum instruction calls, but enables more complex quantum circuit synthesis via \texttt{ctrl}, \texttt{adj}, and \texttt{pow} quantum gate modifiers. Standard \texttt{while} and range-based \texttt{for} loops are also allowed, as well as the conditional \texttt{if-else} block. 

While our work seeks to ensure that our compiler implementation is fully compatible with the base grammar specification for OpenQASM, we also are in a unique position to enhance it with features pertinent to the \texttt{qcor} compiler platform and its user base. We envision the language and compiler presented in this work as a novel language extension for the \texttt{qcor} quantum kernel programming model, i.e. enabling users of \texttt{qcor} to program quantum kernels using our extended OpenQASM language. We seek extensions that (while remaining backwards compatible) enable a more C-like syntax, C-like primitive type declarations, and common quantum programming patterns already present in the \texttt{qcor} language. To start, we have extended the grammar to provide familiar typedefs for 32 and 64 bit integers and floats, Specifically, we parse \texttt{int} as \texttt{int[32]}, \texttt{int64\_t} as \texttt{int[64]}, \texttt{float} as \texttt{float[32]}, and \texttt{double} as \texttt{float[64]}. We have also updated the grammar and implemented the parser to handle both range-based C-like \texttt{for} statements as well as the usual \texttt{for} statement with with initializer, conditional expression, and iteration expression.
\begin{figure}[t]
\begin{subfigure}[b]{\columnwidth}
\caption{Compute-Action ANTLR Grammar Addition.}
\label{fig:ca-grammar}
\vspace{-2mm}
\begin{minted}[frame=single,framesep=2pt, fontsize=\footnotesize, breaklines, escapeinside=||, linenos, numbersep=1mm]{asm}
compute_action_stmt
    : 'compute' compute_block=programBlock 
      'action' action_block=programBlock ;
\end{minted}
\vspace*{2mm}
\end{subfigure}
\begin{subfigure}[b]{\columnwidth}
\caption{OpenQASM code leveraging the compute-action statement.}
\label{fig:ca-grammar-example}
\vspace{-2mm}
\begin{minted}[frame=single,framesep=2pt, fontsize=\footnotesize, breaklines, escapeinside=||, linenos, numbersep=1mm]{llvm}
qubit q[4];
let bottom_three = q[1:3];
compute {
    rx(1.57) q[0];
    h bottom_three;
    for i in [0:3] {
      cnot q[i], q[i + 1];
    }
} action {
    rz(2.2) q[4];
}
\end{minted}
\vspace*{2mm}
\end{subfigure}
\caption{OpenQASM language extension for \texttt{compute}-\texttt{action}-\texttt{uncompute} pattern.}
\end{figure}

Finally, we see an opportunity to enable syntax and semantics for specific compile-time optimizations. We have updated the OpenQASM grammar with support for the ubiquitous \texttt{compute-action-uncompute} pattern, first demonstrated in \cite{projectQ}. Given the common pattern $W = U V U^\dagger$, the new syntax enables one to express $U$ and $V$ as the code in the \texttt{compute} and \texttt{action} scopes, respectively, and the compiler auto-generates the $U^\dagger$ code after application of the \texttt{compute}, \texttt{action} segments. This is demonstrated in Figure \ref{fig:ca-grammar} (addition to OpenQASM grammar) and \ref{fig:ca-grammar-example} (example usage), and is not only useful for readability, but it also gives the compiler implementation the opportunity to generate optimal quantum code in the situation where the programmer wants to program $W$ controlled on the state of another qubit. In this case, the compiler synthesizes $U$ ctrl-$V$ $U^\dagger$ as opposed to the naive ctrl-$U$ ctrl-$V$ ctrl-$U^\dagger$, thereby leading to less multi-qubit operations and shorter depth quantum programs. 

\section{Compiler Architecture}
The architecture we put forward starts with the definition of a frontend parser for the OpenQASM language. This parser produces an AST and we provide custom tree walkers that traverse the tree and construct a corresponding MLIR representation. To handle variable data and scoping, we introduce a custom symbol table, enabling the tracking of variable use-define chains as well as scope visibility. Our MLIR representation is amenable to general transformation, and we leverage this for quantum-level optimizations. Ultimately, our architecture lowers the MLIR down to the LLVM IR. At this level, the generation of executable code and linking with a valid QIR runtime implementation is readily accomplished with standard assemblers and linkers. 
\subsection{Symbol Table}
\label{sec:symbol_table}
The symbol table is the data structure used by the compiler to cache information about each observed symbol (e.g., variable name, its type, its constness, etc.). Since OpenQASM allows for local variables, the symbol table becomes critical for tracking metadata about the variable symbol and its subsequent use. In other words, when processing each statement, the compiler, via the symbol table, is aware of the context of all \emph{visible} symbols, i.e., those from this scope and those above, to perform proper name lookup. Therefore the symbol table provides a mechanism to validate various semantic errors, such as illegal operations for a specific variable type or referring to out-of-scope variables, which could not be detected by syntactic considerations.

We have implemented a symbol table that is composed of an array of scope-indexed hash maps. Each map is a lookup table from variable name to the corresponding MLIR \texttt{mlir::Value} instance representing the variable. Name lookup is performed from the current scope upward (to parent scopes) to find the first match, i.e., the one in the nearest parent scope. Another utility of the symbol table is the compile-time evaluation of constant expressions. OpenQASM supports constant integer and floating-point variable declarations (via the \texttt{const} keyword). The symbol table tracks these constant values and provides a utility to evaluate simple math expressions\footnote{using the C++ Mathematical Expression Toolkit (\texttt{exprtk}) Library} involving these constants at compile time, if possible. 

Critically, the compiler relies on the symbol table to track qubit use-define chains for quantum instruction operations (typical quantum gate invocations). We have designed our quantum instruction operation in the quantum MLIR dialect extension to adhere to the value semantics representation first described in \cite{ittah2021enabling}, whereby quantum operations consume one or many qubit \texttt{mlir::Value} instances and produce one or many new \texttt{mlir::Value} instances as operation return types. Using the underlying pointer to the MLIR variable (\texttt{mlir::Value}) as the lookup key, the symbol table replaces input qubit operands with the newly-created output \texttt{mlir::Value}. Therefore, the OpenQASM source code is compiled into the MLIR representation with explicit use-define chains for qubits amendable to compile-time optimization techniques similar to the DAG representation of quantum circuits. 

\subsection{ANTLR Parser}
\begin{figure}[t!]
\begin{tikzpicture}[node distance=1cm, every node/.style={fill=white, font=\scriptsize}, align=center]
  \node (src) [rectangle, draw=black, minimum width=2cm]{\texttt{Source.qasm}};
  
  \node (lexer) [rectangle, rounded corners, draw=black, minimum width=2cm, minimum height=0.7cm, font=\scriptsize\sffamily, below of=src]{qasm3Lexer};
  
  \node (token_stream) [below of=lexer]{Token Stream};
  
  \node (parser) [rectangle, rounded corners, draw=black, minimum width=2cm, minimum height=0.7cm, font=\scriptsize\sffamily, below of=token_stream]  {qasm3Parser};
  
  \node (AST) [below of=parser]{ANTLR AST};
  
  \node (visitor)[rectangle, rounded corners, draw=black, minimum width=2cm, minimum height=0.7cm, font=\scriptsize\sffamily, below of=AST] {qasm3Visitor};
  
  \node (MLIR) [ellipse, draw=black, fill=blue!20, below of=visitor]{\textbf{MLIR Op Tree}};
    
  \node (grammar) [rectangle, rounded corners, draw=black, minimum width=2cm, left of = lexer, xshift=-1.5cm]{ANTLR grammar \\ \texttt{OpenQASM.g4}};    
  
  \node (error_listener) [rectangle, rounded corners, draw=black, minimum width=2cm, minimum height=0.7cm, font=\scriptsize\sffamily, right of=parser, xshift=1.5cm]  {Error Listener};
  \node (error) [rectangle, draw=black, fill=red!60, font=\scriptsize\sffamily, below of=error_listener]  {\textbf{Syntax Errors!}};
  
  \draw[->] (src) -- (lexer);    
  \draw[->] (lexer) -- (token_stream);
  \draw[->] (token_stream) -- (parser);
  \draw[->] (parser) -- (AST);
  \draw[->] (AST) -- (visitor);
  \draw[->] (visitor) -- (MLIR);
  
  \node[inner sep=0,minimum size=0,left of=lexer, xshift=-0.25cm] (ll) {}; 
  \node[inner sep=0,minimum size=0,left of=parser, xshift=-0.25cm] (pl) {}; 
  \draw[->] (grammar) to (lexer);
  \draw[->] (pl) -- (parser);
  \draw (pl) -- (ll);
  
  \draw[black,dotted,fill=green!20, fill opacity=0.2] ($(grammar.north west)+(-0.2,0.2)$)  rectangle ($(parser.south east)+(0.2,-0.2)$);
  \node [below of=grammar, fill=none, font=\footnotesize] {\textbf{Auto-generated}};
  
  \draw[->] (error_listener) -- (parser);
  \draw[->] (error_listener) -- (error);
\end{tikzpicture}
\caption{ANTLR-based compiler front-end: lexer and parser are auto-generated from the extended OpenQASM grammar (ANTLR grammar file). An error listener is attached to the parser to capture and report any syntax violations encountered during parsing. The \texttt{qasm3Visitor} visits each node of the ANTLR AST constructing the MLIR syntax tree in the standard, affine, and quantum dialects.}
\label{fig:antlr_diagram}
\end{figure}
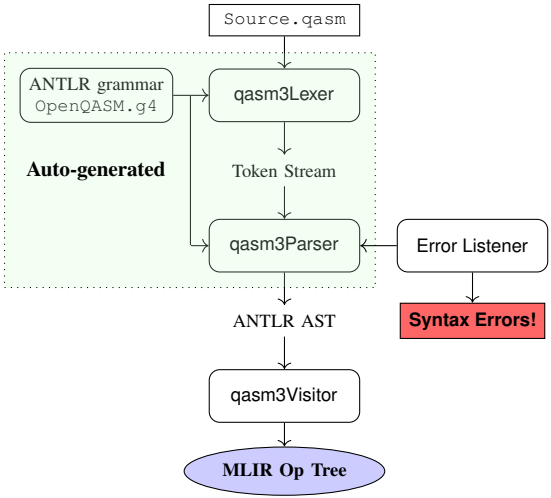

Our compiler implementation leverages the ANTLR~\cite{ANTLR} (\textbf{AN}other \textbf{T}ool for \textbf{L}anguage \textbf{R}ecognition) toolchain to generate the compiler frontend as depicted in Figure \ref{fig:antlr_diagram}. 
With our \emph{extended} OpenQASM grammar as input, ANTLR generates the corresponding lexer and parser utilities capable of scanning and parsing source strings according to the provided grammar rules. The compiler frontend produces an AST representing the input source code against the set of syntactic rules in the grammar. For instance, a valid OpenQASM loop (matching a syntax rule named \texttt{loopStatement}) will be parsed into a \texttt{LoopStatementContext} AST node along with all nested sub-nodes, e.g., the loop termination conditions and the loop body. ANTLR also generates a base AST visitor interface for each grammar file, which includes all possible AST node types that the parser may produce. The AST visitor is the mechanism we leverage to transform the raw OpenQASM syntax tree into the MLIR representation as we will discuss in the next subsection.
  
While processing the input source, the parser may throw exceptions indicating syntactic or semantic errors. The compiler implements the ANTLR \texttt{BaseErrorListener} interface (see Figure~\ref{fig:antlr_diagram}) to catch these potential issues and report them to users with detailed information, such as the location of offending characters.   

\subsection{Visitor Handlers}
Once the valid OpenQASM source has been transformed into the ANTLR AST, the compiler traverses each node in the AST in a depth-first manner producing the equivalent MLIR tree using the standard (operations for classical control flow and memory references), affine (operations for looping), and quantum dialects (our contribution modeling quantum operations).  Table~\ref{table:qasm_mlir} summarized OpenQASM-MLIR rewrite patterns for important OpenQASM constructs.

\begin{table*}[h!]
\caption{OpenQASM to MLIR}
\centering
\begin{tabular}{p{0.15\textwidth}|p{0.25\textwidth} | p{0.55\textwidth}} 
\textbf{Constructs} & \textbf{OpenQASM} & \textbf{MLIR} \\ [0.5ex] 
\hline
\hline
\textbf{Quantum Types}  &  & \\ 
Qubit Register & \texttt{qubit qubit\_array[20];} &  \mintinline{llvm}{
\hline
\textbf{Classical Types} &  & \\ 
Bits & \texttt{bit[20] bit\_array;}  & \mintinline{llvm}{
Integers & \texttt{int[16] short\_int;}  & \mintinline{llvm}{
Floating point numbers & \texttt{float[32] sp\_float;}  & \mintinline{llvm}{
Global Constants & \texttt{const shots = 1024;}  & \mintinline{llvm}{global_memref "private" constant @shots : memref<i64> = dense<1024>} \\ 
\hline
\textbf{Quantum Instructions} &   &  \\ [0.5ex] 
Gates & \texttt{ry(theta) q;}  & \mintinline{llvm}{
 &   & Note: \texttt{\%2: !quantum.Qubit}; \texttt{\%3: f64}   \\ [0.5ex] 
Measurements & \texttt{b = measure q;}  & 
\vspace{-2mm}
\begin{minted}{llvm}
%7 = q.mz(%6) : !quantum.Result
%9 = q.resultCast(%7) : i1
store %9, %4[] : memref<i1>
\end{minted} 
\\[-3mm]
&   & Note: \texttt{\%6: !quantum.Qubit}; \texttt{\%4: memref<i1>}  \\
\hline
\textbf{Classical Operations} & \texttt{a += 4;}  & 
\vspace{-2mm}
\begin{minted}{llvm}
%c4_i64 = constant 4 : i64
%1 = load %0[] : memref<i64>
%2 = addi %1, %c4_i64 : i64
store %2, %0[] : memref<i64>
\end{minted} 
\\ [-3mm]
&   & Note: \texttt{\%0: memref<i64>} (represents \texttt{a} variable)  \\ [0.5ex] 
\hline
\textbf{Branching} & \texttt{if (i == 5) \{...\}}  & 
\vspace{-2mm}
\begin{minted}{llvm}
  %c5_i64 = constant 5 : i64
  %1 = load %0[] : memref<i64>
  %2 = cmpi "eq", %1, %c5_i64 : i64
  cond_br %2, ^bb1, ^bb2
^bb1:  // pred: ^bb0
  br ^bb2
^bb2: // 2 preds: ^bb0, ^bb1
\end{minted} 
\\ [-3mm]
&   & Note: \texttt{\%0: memref<i64>} (represents \texttt{i} variable)  \\ [0.5ex] 
\hline
\textbf{Looping} & \texttt{for i in [0:10] \{...\}}  & 
\vspace{-2mm}
\begin{minted}[breaklines]{llvm} 
affine.for %arg0 = affine_map<(d0) -> (d0)>(%1) to affine_map<(d0) -> (d0)>(%0) {...}
\end{minted} 
\\
&   & Note: \texttt{\%0: index} and \texttt{\%1: index} represent the constant values of 10 and 0, respectively. \\
\hline
\textbf{Subroutines} & \texttt{def foo(float[64]:theta) qubit[2]:q \{...\}}  & 
\vspace{-2mm}
\begin{minted}{llvm}
func @foo(%arg0: f64, %arg1: !quantum.Array) {
    ...
    return
}
\end{minted}
\\ 
 & \texttt{foo(theta) qq;}  & \mintinline{llvm}{call @foo(
 &   & Note: \texttt{\%2: f64} and \texttt{\%0: !quantum.Array} represent the \texttt{theta} variable and the \texttt{q} qubit array, respectively. \\ [0.5ex] 
\hline
\textbf{Extern functions} & \texttt{extern foo(float[64]) -> float[64];}  & \mintinline{llvm}{func private @foo(f64) -> f64} \\
\hline
\textbf{Modifiers} &   &  \\ 
Adjoint & \texttt{inv @ phase(pi) q;}  & 
\vspace{-3mm}
\begin{minted}{llvm}
q.adj_region {
    %2 = qvs.phase(%1, %cst) : !quantum.Qubit
} 
\end{minted}
\\ 
Controlled & \texttt{ctrl @ oracle q[0], q[1];}  & 
\vspace{-3mm}
\begin{minted}{llvm}
q.ctrl_region {
 %3 = call @oracle(%2) : (!quantum.Qubit) -> !quantum.Qubit
} (ctrl_bit = %1) 
\end{minted}
\\ 
Power & \texttt{pow(8) @ foo q;}  & 
\vspace{-3mm}
\begin{minted}{llvm}
q.pow_u_region {
    %c8_i64 = constant 8 : i64
    %2 = call @foo(%1) : (!quantum.Qubit) -> !quantum.Qubit
} (pow = %c8_i64)
\end{minted}
\\ 
\hline
\textbf{Aliasing} &  &  \\ [0.5ex] 
Slicing & \texttt{let slice = reg[0:2:12];}  & 
\vspace{-3mm}
\begin{minted}[breaklines]{llvm}
%1 = q.qarray_slice(%0, %c0_i64, %c2_i64, %c12_i64) : !quantum.Array
\end{minted}
\\ [0.5ex] 
Concatenation & \texttt{let concat = reg1 || reg2;}  & 
\vspace{-3mm}
\begin{minted}[breaklines]{llvm}
%2 = q.qarray_concat (%0, %1) : !quantum.Array
\end{minted}
\\ [0.5ex] 
\hline
\end{tabular}
\label{table:qasm_mlir}
\end{table*}

In particular, quantum types (\texttt{qubit} and \texttt{qreg}) and classical types (boolean, variable-width integer or floating-point numbers, and arrays) are mapped to QIR types (\texttt{Qubit} and \texttt{Array}) or memory-referenced (\texttt{memref}) MLIR Standard dialect types (e.g., \texttt{i1} for boolean bits, \texttt{i8} for 8-bit integers, etc.) Classical math operations are converted to the corresponding instructions from the MLIR Standard dialect, such as \texttt{addi} or \texttt{cmpi} for integer addition or comparison, respectively. Importantly, OpenQASM \texttt{for} loops are transformed into an \texttt{AffineForOp}~\cite{mlir-affine} (MLIR affine dialect) amenable to future classical optimization passes, such as loop unrolling.

Intrinsic quantum gates are converted to value-semantics quantum operations (static single assignment form) of the quantum dialect, as shown in Table~\ref{table:qasm_mlir}. As described in Sec.~\ref{sec:symbol_table}, we use the symbol table to track the qubit operands (as opaque \texttt{mlir::Value} pointers) and replace them with the new values created by each value-semantics quantum gate operation. In other words, each qubit SSA variable (shown as \texttt{\%k} in Table~\ref{table:qasm_mlir}) will only be assigned and used once, thus allowing us to trace gate operations on each qubit line. This is to explicitly define the use-define chains, which we can leverage in downstream quantum optimizations.

Another key feature of OpenQASM is the ability to express quantum gate modifiers (e.g., controlled or adjoint) for both intrinsic gates and subroutines. Our compiler implementation takes a pragmatic approach by rewriting modifiers into scoped regions with dedicated MLIR \emph{marker} operations as shown in Table~\ref{table:qasm_mlir}. These operations are effectively \texttt{no-ops}, but indicate to the runtime that the following region of quantum operations is to be handled differently. For example, for the \texttt{ctrl} marker (\texttt{q.ctrl\_region}), the operations within that region should be processed to synthesize the controlled version of that composite operation. We preserve the high-level semantics of these modifiers at both the MLIR and latter LLVM IR levels (see~\ref{sec:quantum_to_llvm}) rather than trying to perform compile-time gate synthesis. Ultimately, the evaluation and synthesis of these modifier-enclosed blocks will be performed by QIR-compatible runtime implementation. 

To handle the nested, recursive nature of the OpenQASM syntax tree, the compiler uses a multi-layer visiting strategy whereby a standalone visitor-like utility, named \texttt{qasm3\_expression\_generator}, is provided to traverse and process sub-expression nodes in-place, if necessary. To give an example, when visiting a \texttt{for} loop with a math expression as its upper bound, the main AST visitor would use this \texttt{qasm3\_expression\_generator} to handle this sub-expression (converting the math expression to a MLIR equivalent) and then take the resulting value (as a \texttt{mlir::Value}) to construct the current MLIR \texttt{for} loop.

\subsection{Progressive Lowering}
\begin{figure}[t!]
\begin{tikzpicture}[node distance=1.5cm, every node/.style={fill=white, font=\scriptsize}, align=center, ->, >=stealth', level/.style={sibling distance = 1cm/#1, level distance = 0.75cm},transform shape]
  \node (src) [rectangle, rounded corners, draw=black, minimum width=1cm, , minimum height=0.7cm]{OpenQASM};
  \node (antlr_ast)[rectangle, right of=src, minimum width=1cm] {};
  \node (antlr_ast_plot)[right of=src, yshift=0.75cm, circle, fill=red!60, align=center, minimum size=0.1cm] {}
    child
    {
        node [circle, fill=red!60, align=center, minimum size=0.1cm] {} 
        child
        {
            node [circle, fill=red!60, align=center, minimum size=0.1cm] {} 
        }
        child
        {
            node [circle, fill=red!60, align=center, minimum size=0.1cm] {} 
        }
    };
  \node (antlr_text) [rectangle, rounded corners, minimum width=1cm, below of=antlr_ast_plot, yshift=-0.65cm]{ANTLR\\AST};
  \node (mlir) [rectangle, rounded corners, draw=black, minimum width=1cm, minimum height=0.7cm, right of=antlr_ast]{MLIR};
  \node (mlir_llvm) [rectangle, rounded corners, draw=black, minimum width=1cm, right of=mlir]{MLIR\\LLVM};
  \node (llvm) [rectangle, rounded corners, draw=black, minimum width=1cm, minimum height=0.7cm, right of=mlir_llvm]{LLVM\\IR};
  \node (exe) [rectangle, rounded corners, draw=black, minimum width=1cm, minimum height=0.7cm, right of=llvm]{Binary\\Exe};

  \draw[->] (src) -- (antlr_ast);
  \draw[->] (antlr_ast) -- (mlir);
  \path[draw, ->] (mlir) -- node [midway,above,text opacity=1, fill opacity=0,yshift=0.05cm]{\tiny Lower} (mlir_llvm);
  \path[draw, ->] (mlir_llvm) -- node [midway,above,text opacity=1, fill opacity=0,yshift=0.05cm]{\tiny\texttt{-O3}} (llvm);
  \draw[->] (llvm) -- (exe); 
  
  \draw[->] (mlir.north east)  to [looseness=3]  node[above]{Optimization} (mlir.north west);
     
  \node (mlir_text1) [anchor=west, text width=1.5cm, below of=mlir, yshift=0.95cm]{\tiny{Quantum Dialect}};
  \node (mlir_text2) [anchor=west, text width=1.5cm, below of=mlir, yshift=0.65cm, xshift=-0.1cm]{\tiny{Affine Dialect}};
  \node (mlir_text3) [anchor=west, text width=1.5cm, below of=mlir, yshift=0.35cm]{\tiny{Standard Dialect}};
  
  \node (mlir_llvm_text) [anchor=west, text width=1.5cm, below of=mlir_llvm, yshift=0.95cm]{\tiny{LLVM Dialect}};
  \node (llvm_text) [anchor=west, text width=1.5cm, below of=llvm, yshift=0.95cm]{\tiny{QIR + LLVM}};
    
\end{tikzpicture}
\caption{Compilation pipeline: The ANTLR-based frontend parses the OpenQASM source string into an Abstract Syntax Tree (AST) data structure. By processing (visiting) the AST, we generate a MLIR representation using a couple of dialects, most importantly, our quantum dialect. A set of optimization passes can be applied at this stage to simplify the MLIR tree before it is lowered to the LLVM dialect. At this stage, the IR tree only contains valid LLVM instructions including QIR-adherent function calls and types. Standard LLVM optimization can be applied when the LLVM dialect is lowered to bitcode, e.g., the \texttt{-O3} LLVM optimization flag. Lastly, the LLVM IR bitcode is compiled to binary executable by linking in a compatible QIR runtime implementation.}
\label{fig:compiler_pipeline}
\end{figure}
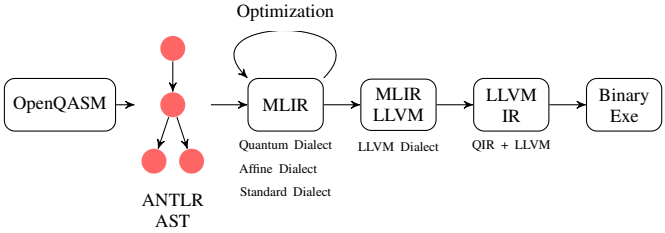

After visiting all the ANTLR AST nodes representing the input OpenQASM program, the compiler has constructed a MLIR code in the quantum, affine, standard, and built-in dialects. As depicted in Figure~\ref{fig:compiler_pipeline}, this is the first stage of a progressive, multi-stage IR transformation and lowering pipeline that produces an optimized executable.

Next, we perform a set of optimization passes at the MLIR level whereby control flow constructs (e.g., those from the affine dialect) and the static single assignment (SSA) form of quantum instructions in the quantum dialect are suitable for static optimization procedures. The optimized MLIR code is lowered to the LLVM dialect via the MLIR \texttt{ConversionPattern} utility. At this point, all quantum-related operations have been lowered to QIR functions. The final lowering to LLVM IR and any built-in LLVM optimizations (e.g., \texttt{-O3} optimization) are provided by the MLIR-LLVM infrastructure, producing binary executables targeting the QIR runtime along with the classical compute ISA (e.g., x86/Arm/OpenPC depending on the target platform).

\subsubsection{Optimization Passes}
\label{sec:opt_passes}
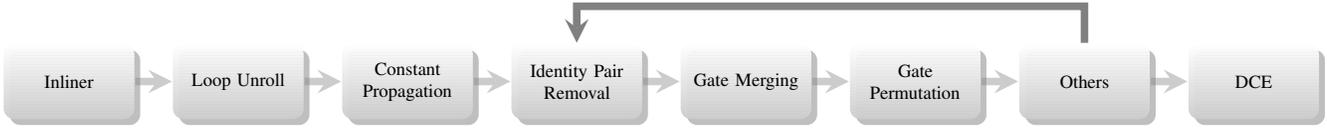
\begin{figure*}[h!]
\smartdiagramset{
 border color=none, 
 module x sep=2.25, 
 back arrow disabled, 
 font=\scriptsize,
 module minimum width=0.8cm,
 uniform color list=gray!40 for 8 items,
}
\smartdiagramadd[flow diagram:horizontal]{Inliner, Loop Unroll, Constant Propagation, Identity Pair Removal, Gate Merging, Gate Permutation, Others, DCE}{}
\begin{tikzpicture}[overlay]
\draw[additional item arrow type] (module7) -- ++(0,1) -| (module4);
\end{tikzpicture}
\caption{Optimization pass pipeline. Optimization passes simplifying quantum gate operations are repeated a set number of times.}
\label{fig:opt_pipeline}
\end{figure*}

Figure~\ref{fig:opt_pipeline} illustrates the MLIR-level optimization pipeline that we have implemented in the OpenQASM compiler. 
Specifically, we combine optimization techniques from both classical and quantum programming, such as function inlining, loop unrolling, and various quantum circuit optimization procedures. Table~\ref{table:mlir_pass_desc} lists the quantum optimization passes that we have implemented for the MLIR operations in our quantum dialect. Inlining and loop unrolling are built-in MLIR passes for the standard and affine dialects, respectively.

\begin{table}[h!]
\caption{List of MLIR Passes}
\centering
\begin{tabular}{||p{0.2\columnwidth}|p{0.7\columnwidth}||} 
 \hline
 Pass Name & Descriptions \\ [0.5ex] 
 \hline\hline
 Identity Pair Removal & Simplify or remove redundant quantum instructions. For example, this pass removes any gates immediately followed by their adjoints, such as pairs of $X$-$X$, $T$-$T^\dagger$, or $CNOT$-$CNOT$ gates on the same qubits. Repeated qubit reset instructions are also simplified.  \\ 
 \hline
 Rotation Merging & Combine consecutive mergable quantum instructions, e.g., $Z$ and $Rz(\theta)$, $Rx(\theta_1)$ and $Rx(\theta_2)$, etc.  \\
 \hline
 Gate Sequence Simplification & Find a sequence of consecutive compile-time constant gates and simplify if possible, i.e., resynthesize to fewer gates. For example, $H$-$T$-$H$ gate sequence can be simplified to $Rx(\pi/4)$.   \\
 \hline
 Qubit Extract Lifting & Merge duplicate qubit extracts from registers with compile-time constant indices. This pass also unifies the SSA use-define chain after loop unrolling (loop induction variable as qubit array index) and function inlining. \\
 \hline
 Gate Permutation & Permute gates that are commutative, e.g., $Rz$ on the control qubit of a $CNOT$ gate. Despite no immediate benefit (no gate count reduction), this pass might produce optimization opportunities for others, such as rotation merging.  \\
 \hline
 Constant Propagation & Propagate global constants, e.g., constant integer values as loop counts or constant floating points as rotation angles.  \\
 \hline
 Dead Code Elimination (DCE) & Eliminate unused operations (dead code). For example, \texttt{qreg} allocation whose result is never used can be eliminated. These dead values may emerge as a result of other passes.   \\ [1ex] 
 \hline
\end{tabular}
\label{table:mlir_pass_desc}
\end{table}

The pseudocode in Algorithm~\ref{alg:gate_merging} illustrates a typical MLIR optimization pass based on dataflow analysis. Specifically, we show the procedure to perform quantum gate merging on the MLIR AST tree. By adhering to value-semantics for the quantum operations, we are able to follow the use-define chain of each quantum instruction (shown as the \texttt{User} map in Algorithm~\ref{alg:gate_merging}). With the SSA dataflow information, we can query the next quantum operation on the qubit line and check whether a gate-merge opportunity exists. For illustration purposes, we depict the gate merging procedure as two black-box functions, \texttt{CanMerge} and \texttt{Merge}, implementing checking and gate generation procedures. The new injected gate operation will have its input and output SSA values bridging those two original instructions (line 10 and 11 in Algorithm~\ref{alg:gate_merging}). Each optimization pass operating on the MLIR operation tree maintains the SSA value chain as they transform the IR.

\begin{algorithm}\captionsetup{labelfont={sc,bf}}
\caption{Gate Merging Optimization}
\textbf{Vars}: 
\begin{itemize}
    \item \texttt{ops} : [VSOp] (Sequence of value semantics ops)
    \item \texttt{Users}: VSOp $\mapsto$ [VSOp] (use-define trace mechanism)
    \item \texttt{CanMerge}: (VSOp, VSOp)  $\mapsto$ $\mathbb{B}$ (Mergeable check)
    \item \texttt{Merge}: (VSOp, VSOp)  $\mapsto$ VSOp (Create merge op) 
\end{itemize}
\textbf{Gate Merging}:  
\begin{algorithmic}[1]
\STATE{\texttt{dead\_ops}: [VSOp] $\gets$ \texttt{[]}}
\FOR{\texttt{op} $\in$ \texttt{ops}} 
    \IF{\texttt{op} $\in$ \texttt{dead\_ops}} 
        \STATE{Skip to next op} 
    \ENDIF
    \IF{\texttt{Length(Users(op)) == 1}} 
        \STATE{\texttt{next\_op} $\gets$ \texttt{Users(op)[0]}}
        \IF{\texttt{CanMerge(op, next\_op)}} 
            \STATE{\texttt{merged\_op} $\gets$ \texttt{Merge(op, next\_op)}}
            \STATE{\texttt{merged\_op.input} $\gets$ \texttt{op.input}}
            \STATE{\texttt{merged\_op.output} $\gets$ \texttt{next\_op.output}}
            \STATE{Add \texttt{merged\_op} to IR tree}
            \STATE{Append \texttt{op} and \texttt{next\_op} to \texttt{dead\_ops}}
        \ENDIF
    \ENDIF
\ENDFOR
\FOR{\texttt{op} $\in$ \texttt{dead\_ops}} 
    \STATE{Erase \texttt{op} from IR tree}
\ENDFOR
\end{algorithmic}
\label{alg:gate_merging}
\end{algorithm}

We also want to note that this optimization procedure, as well as others listed in Table~\ref{table:mlir_pass_desc}, is most effective when the AST is a flat linear region whereby the use-define chain is uninterrupted (e.g., due to subroutine calls or loops). Therefore, it is crucial to have loop unrolling and function inlining passes applied beforehand as shown in Figure~\ref{fig:opt_pipeline}. In this pass pipeline, some passes, especially those performing quantum circuit optimization, are applied multiple times in a loop to make sure that we can pick up new optimizing patterns that emerged thanks to the code rewrite of previous passes.

In Figure~\ref{fig:opt_example}, we demonstrate the MLIR transformation along the optimization pipeline for a simple OpenQASM source code (Figure~\ref{fig:opt_example:qasm}).
\begin{figure}[t]
\begin{subfigure}[b]{\columnwidth}
\caption{OpenQASM source}
\label{fig:opt_example:qasm}
\vspace{-2mm}
\begin{minted}[frame=single,framesep=2pt, fontsize=\footnotesize, breaklines, escapeinside=||, linenos, numbersep=1mm]{asm}
def foo qubit[2]:qq {
  cx qq[0], qq[1];
}

qubit q[2];
foo q;
cx q[0], q[1];
\end{minted}
\vspace*{2mm}
\end{subfigure}

\begin{subfigure}[b]{\columnwidth}
\caption{Unoptimized MLIR}
\label{fig:opt_example:mlir}
\vspace{-2mm}
\begin{minted}[frame=single,framesep=2pt, fontsize=\footnotesize, breaklines, escapeinside=||, linenos, numbersep=1mm]{llvm}
%0 = q.qalloc(2) { name = q } : !quantum.Array
call @foo(%0) : (!quantum.Array) -> ()
%c0_i64 = constant 0 : i64
%1 = q.extract(%0, %c0_i64) : !quantum.Qubit
%c1_i64 = constant 1 : i64
%2 = q.extract(%0, %c1_i64) : !quantum.Qubit
%3:2 = qvs.cx(%1, %2) : !quantum.Qubit, !quantum.Qubit
q.dealloc(%0)
return
\end{minted}
\vspace*{2mm}
\end{subfigure}

\begin{subfigure}[b]{\columnwidth}
\caption{MLIR after inlining pass}
\label{fig:opt_example:mlir_after_inline}
\vspace{-2mm}
\begin{minted}[frame=single,framesep=2pt, fontsize=\footnotesize, breaklines, escapeinside=||, linenos, numbersep=1mm]{llvm}
%c0_i64 = constant 0 : i64
%c1_i64 = constant 1 : i64
%0 = q.qalloc(2) { name = q } : !quantum.Array
%1 = q.extract(%0, %c0_i64) : !quantum.Qubit
%2 = q.extract(%0, %c1_i64) : !quantum.Qubit
%3:2 = qvs.cx(%1, %2) : !quantum.Qubit, !quantum.Qubit
%6:2 = qvs.cx(%3#0, %3#1) : !quantum.Qubit, !quantum.Qubit
q.dealloc(%0)
return
\end{minted}
\vspace*{2mm}
\end{subfigure}

\begin{subfigure}[b]{\columnwidth}
\caption{MLIR after identity pair removal pass}
\label{fig:opt_example:mlir_after_id_remove}
\vspace{-2mm}
\begin{minted}[frame=single,framesep=2pt, fontsize=\footnotesize, breaklines, escapeinside=||, linenos, numbersep=1mm]{llvm}
%c0_i64 = constant 0 : i64
%c1_i64 = constant 1 : i64
%0 = q.qalloc(2) { name = q } : !quantum.Array
%1 = q.extract(%0, %c0_i64) : !quantum.Qubit
%2 = q.extract(%0, %c1_i64) : !quantum.Qubit
q.dealloc(%0)
return
\end{minted}
\vspace*{2mm}
\end{subfigure}

\begin{subfigure}[b]{\columnwidth}
\caption{MLIR after DCE}
\label{fig:opt_example:mlir_final}
\vspace{-2mm}
\begin{minted}[frame=single,framesep=2pt, fontsize=\footnotesize, breaklines, escapeinside=||, linenos, numbersep=1mm]{llvm}
return
\end{minted}
\vspace*{2mm}
\end{subfigure}

\caption{MLIR optimization example. The input OpenQASM source code (a) is first compiled into the MLIR representation (b), which is then processed by a sequence of optimization passes. After the call to \texttt{foo} (b, line 2) is inlined (c), the back-to-back CNOT pattern emerges; thus, both gates are removed by the identity pair removal pass (d). Finally, the DCE pass eliminates the redundant qubit array allocation, constant value declarations, and qubit extract calls (e) since they have no further use.}
\label{fig:opt_example}
\end{figure}
This code contains a subroutine definition and later invocation, which is compiled to the MLIR \texttt{CallOp} (line 2 in Figure~\ref{fig:opt_example:mlir}). This call is then inlined (Figure~\ref{fig:opt_example:mlir_after_inline}, line 4-6), resulting in a CNOT-CNOT identity pair which is removed by the identity pair removal pass (Figure~\ref{fig:opt_example:mlir_after_id_remove}). What is left after this step is a sequence of unused operations, such as extracting qubit addresses and the \texttt{qreg} allocation itself. These are all dead code, hence removed by the final DCE pass as shown in Figure~\ref{fig:opt_example:mlir_final}.

\subsubsection{Dialect Conversion and Lowering}
\label{sec:quantum_to_llvm}
After simplifying the MLIR tree with optimization passes such as those listed in Table~\ref{table:mlir_pass_desc}, the compiler will lower the MLIR representation to LLVM progressively, as depicted in Figure~\ref{fig:compiler_pipeline}. This lowering procedure is similar to the one described in~\cite{mccaskey2021mlir} for OpenQASM 2 compilation. We implemented a collection of \texttt{mlir::ConversionPattern}s to perform the conversion from quantum dialect to the LLVM dialect targeting the QIR specification. 

As compared to the work in~\cite{mccaskey2021mlir}, the lowering pipeline of the \texttt{qcor} compiler has been enhanced with (1) dialect conversion from affine to LLVM branch-based CFG (control flow graph) representation and (2) conversion pattern implementations for new quantum dialect operations. The first one stems from the fact that we utilize operations from the affine dialect to handle control flows (e.g., for loops) in OpenQASM. The latter involves lowering procedures for new quantum dialect operations for gate modifiers (see Table~\ref{table:qasm_mlir}) as well as the new quantum value semantics instruction. MLIR modifier-marked regions (\texttt{ctrl}, \texttt{inv}, or \texttt{pow}) are converted to the calls to corresponding quantum runtime functions at the beginning and the end of the scoped block. To lower the value-semantics quantum operations (MLIR quantum dialect) to memory-semantics LLVM QIR calls, the qubit SSA variables are mapped back to their root variable (in the use-define chain) by propagating the \texttt{mlir::Value} of input qubit operands to the corresponding outputs.

\subsection{QIR Implementation and Linking}

The last stage of the compilation workflow, as shown in Figure~\ref{fig:compiler_pipeline}, involves the compilation of LLVM IR bytecode into a binary object containing QIR function calls, that needs to be linked to a valid QIR runtime implementation to form an executable. As described in~\cite{mccaskey2021mlir}, \texttt{qcor} provides a QIR runtime library implementation backed by the XACC framework~\cite{mccaskey2020xacc}. 

Since quantum value-semantics operations are converted to memory-semantics function calls (e.g., \texttt{void \_\_quantum\_\_qis\_\_INSTNAME(Qubit*,...)}) during the lowering stage, they are compatible with the existing QIR intrinsic quantum gates in the runtime. Key extensions to the QIR runtime to support OpenQASM are the region marker functions to implement gate modifier concepts, such as controlled (\texttt{ctrl}) or adjoint (\texttt{inv}). Specifically, when a quantum gate or subroutine is subjected to a modifier directive, the compiler injects the corresponding runtime functions before and after the modified operation. For example, \texttt{\_\_quantum\_\_rt\_\_start\_adj\_u\_region} and \texttt{\_\_quantum\_\_rt\_\_end\_adj\_u\_region} are functions to denote a region of code whereby the inverse (adjoint) of the collected quantum sequence generated within should be applied. Once again, we take a pragmatic approach in implementing the gate modifier feature of the OpenQASM language by delegating the modified circuit realization to the runtime. By invoking the wrapped code region at runtime in a special instruction collection mode, we can retrieve the flattened sequence of gates and thus construct the corresponding modified circuit (e.g., adjoint or controlled) for backend execution.
\begin{figure}[t]
\begin{minted}[frame=single,framesep=2pt, fontsize=\footnotesize, breaklines, escapeinside=||, linenos, numbersep=1mm]{asm}

OPENQASM 3;
include "stdgates.inc";

const shots = 1024;
// State-preparation:
def ansatz(float[64]:theta) qubit[2]:q {
    x q[0];
    ry(theta) q[1];
    cx q[1], q[0];
}

def compute(float[64]:theta) qubit[2]:q -> float[64] {
    bit first, second;
    float[64] num_parity_ones = 0.0;
    float[64] result;
    for i in [0:shots] {
        ansatz(theta) q;
        // Change measurement basis
        h q;
        // Measure
        first = measure q[0];
        second = measure q[1];
        if (first != second) {
            num_parity_ones += 1.0;
        }
        // Reset
        reset q;
    }

    // Compute expectation value
    result = (shots - num_parity_ones) / shots - num_parity_ones / shots;
    return result;
}

float[64] theta, exp_val;
qubit qq[2];
// Try a theta value:
theta = 0.123;
exp_val = compute(theta) qq;
print("Avg <X0X1> = ", exp_val);
\end{minted}
\caption{OpenQASM example: Compute Pauli expectation ($\sigma_X\sigma_X$) after a state-preparation circuit (\texttt{ansatz}).}
\label{fig:demo_deuteron}
\end{figure}

Finally, the QIR runtime environment can be further specialized via compilation flags or executable invocation arguments. For instance, specific hardware or simulator backend (\texttt{qpu}) can be selected, and the quantum runtime can be configured to run in a tightly-coupled execution mode (simulator-only) whereby the dynamical measurement-controlled branching is fully supported.  

\section{Demonstration}
\label{sec:demonstration}
In this section, we demonstrate the utility and performance of an MLIR-based compiler. We present a typical OpenQASM programming, compilation, and execution workflow using the \texttt{qcor} infrastructure. The compilation speed is benchmarked against a variety of comparable quantum compilers. Lastly, we provide an example showing extensions to the OpenQASM language provided by \texttt{qcor}.
\subsection{Compilation and execution workflow}
\begin{figure}[t]
\begin{minted}[frame=single,framesep=2pt, fontsize=\footnotesize, breaklines, escapeinside=||, linenos, numbersep=1mm]{llvm}
func @compute(%arg0: f64, %arg1: !quantum.Array) -> f64 {
    ...
    br ^bb1
  ^bb1:  // 2 preds: ^bb0, ^bb5
    %5 = load %4[] : memref<i64>
    %6 = cmpi "slt", %5, %c1024_i64 : i64
    cond_br %6, ^bb2, ^bb3
  ^bb2:  // pred: ^bb1
    %7 = q.extract(%arg1, %c0_i64) : !quantum.Qubit
    %8 = qvs.x(%7) : !quantum.Qubit
    %9 = q.extract(%arg1, %c1_i64) : !quantum.Qubit
    %10 = qvs.ry(%9, %arg0) : !quantum.Qubit
    %11 = q.extract(%arg1, %c1_i64) : !quantum.Qubit
    %12:2 = qvs.cx(%11, %8) : !quantum.Qubit, !quantum.Qubit
    ....
    %23 = cmpi "ne", %21, %22 : i1
    cond_br %23, ^bb4, ^bb5
  ^bb3:  // pred: ^bb1
    ...
    %28 = subf %27, %26 : f64
    ...
    %37 = divf %34, %36 : f64
    ...
    return %39 : f64
  ^bb4:  // pred: ^bb2
    %40 = load %2[] : memref<f64>
    %41 = addf %40, %cst_0 : f64
    store %41, %2[] : memref<f64>
    br ^bb5
  ^bb5:  // 2 preds: ^bb2, ^bb4
    %42 = qvs.reset(%14) : !quantum.Qubit
    %43 = qvs.reset(%16) : !quantum.Qubit
    %44 = load %4[] : memref<i64>
    %45 = addi %44, %c1_i64 : i64
    store %45, %4[] : memref<i64>
    br ^bb1
\end{minted}
\caption{Truncated MLIR representation of the OpenQASM program in Figure~\ref{fig:demo_deuteron}. Here, we only show a simplified MLIR printout of the core \texttt{deuteron} function whereby most of the code has been omitted for clarity. High-level classical control flow constructs such as the for loop and the if statement in Figure~\ref{fig:demo_deuteron} are converted into LLVM-style CFG constructs and operations such as blocks and branches. Thanks to MLIR-level optimization (sec.~\ref{sec:opt_passes}), the \texttt{ansatz} subroutine in Figure~\ref{fig:demo_deuteron} has been inlined into the \texttt{deuteron}  body as shown as \texttt{qvs.x}, \texttt{qvs.ry}, and \texttt{qvs.cx} operations. Arithmetic operations  are translated to MLIR operations from the Standard dialect, such as \texttt{addf}, \texttt{subf}, \texttt{divf} (floating-point numbers) or \texttt{addi}, \texttt{cmpi} (integers). }
\label{fig:demo_deuteron_mlir}
\end{figure}
\label{sec:demo_deuteron}

Figure~\ref{fig:demo_deuteron} shows an OpenQASM source code to compute the expectation value of Pauli operators (e.g., $\sigma_{X}\sigma_{X}$ in this case) in an arbitrary state prepared by a variational ansatz. The state-preparation is expressed as a quantum subroutine (\texttt{ansatz}, lines 6-10). This is a prototypical procedure commonly used in the variational quantum eigensolver (VQE) algorithm. 

A feature that we want to highlight is the fact that Pauli expectation accumulation is explicitly expressed as a for loop (lines 16-28). Note the \texttt{h} quantum instruction broadcasts across all qubits in the register. In each iteration, we count the parity of measured bits to compute the average result across all runs (shots) in line 31. An abbreviated MLIR representation of the \texttt{compute} subroutine in Figure~\ref{fig:demo_deuteron} is depicted in Figure~\ref{fig:demo_deuteron_mlir} after all MLIR-level optimization passes have been applied. For the sake of presentation, we only keep the high-level structure of the MLIR printout (omitted regions are presented as ellipses).  

The semantics of the OpenQASM source code is faithfully translated into the MLIR representation consisting of operations from our quantum dialect (e.g., quantum gates) and the standard dialect (e.g., branching and arithmetic instructions). We can now transform this high-level IR to executable code adhering to the QIR specification (see Figure~\ref{fig:compiler_pipeline}). At the time of writing, the execution of this type of tightly coupled quantum-classical program, whereby measurement feedforward is required, is only applicable to simulator backends of \texttt{qcor}'s \texttt{ftqc} runtime~\cite{qcor-acm}. We anticipate that quantum hardware providers will support this dynamical runtime model in the near future as OpenQASM becomes more mature and widely-adopted. Importantly, in our workflow, the runtime implementation is linked in at the final phase of the compilation pipeline, thus could be provided interchangeably and dynamically to target different accelerator targets.

\subsection{Compiler performance}
\label{sec:benchmark}
In this section, we benchmark\footnote{Set-up: Intel Xeon CPU E5-2698 v4 @ 2.20GHz running Linux Debian 10 distribution.} the compilation and resource estimation runtime of the OpenQASM compiler against a set of different quantum programming languages and frameworks, such Q\#\footnote{Microsoft Quantum Development Kit 0.18.2107153439}, Qiskit\footnote{qiskit-terra 0.18.1}, and t$|$ket$\rangle$\footnote{pytket 0.13.0}. We benchmark the total runtime required to generate and execute binary executables in the resource estimation mode, i.e., counting flattened quantum gates, because it provides a mechanism to compare statically-compiled executables against Python-based interpreted scripts constructing the equivalent circuits.

\begin{figure}[t!]
\centering
\begin{tikzpicture}[every mark/.append style={mark size=1pt}]
\begin{semilogyaxis}[
      legend columns=3,
      legend style={at={(1.0,1.1)},anchor=north east, nodes={scale=0.745, transform shape}},
      xmin = 5, xmax = 50,
      xlabel = {N  qubits}, 
      ylabel = {Runtime [secs]}, 
      y label style={at={(axis description cs:0.1, 0.55)},anchor=south}]
\addplot+ [error bars/.cd, 
    y fixed,
    y dir=both, 
    y explicit]
table[x=x, y=y,y error=error] {
x  y                        error
5	14.553841972351075      1.2926747249971386 
10	46.19267573356628       2.641008650762535
20	191.8102480173111       2.4560518751452363 
30	457.4762924194336       18.038306222883786
40	871.0876515865326       13.857947978864567
50	1476.1323881864548      41.27279442136586
};
\addlegendentry{Qiskit};

\addplot+ [error bars/.cd, 
    y fixed,
    y dir=both, 
    y explicit]
table[x=x, y=y,y error=error] {
x  y                    error
5	2.4804011821746825  0.07056878106402498
10	7.892903161048889   0.4176733572162844
20	33.78146722316742   1.5680721861532423
30	82.36961619853973   1.1738009416886714
40	140.1442172050476   2.0458854405760603
50	224.45344417095185  3.2545876946712133
};
\addlegendentry{t$|$ket$\rangle$};

\addplot+ [error bars/.cd, 
    y fixed,
    y dir=both, 
    y explicit]
table[x=x, y=y,y error=error] {
x  y                        error
5	17.15200607776642       0.40707767778831905
10	16.632620406150817      0.7878573308104417
20	16.638732981681823      0.22162641965199445
30	16.457424235343932      0.48298625856473215
40	16.700783801078796      0.31651494997119584
50	16.814245843887328      0.2746374910642685
};
\addlegendentry{Q\#-compile};

\addplot+ [error bars/.cd, 
    y fixed,
    y dir=both, 
    y explicit]
table[x=x, y=y,y error=error] {
x  y            error
5	1.107       0.025
10	1.149       0.038
20	1.189       0.032
30	1.244       0.022
40	1.288       0.058
50	1.283       0.099
};
\addlegendentry{OpenQASM-compile};

\addplot+ [green!50!black, error bars/.cd, 
    y fixed,
    y dir=both, 
    y explicit]
table[x=x, y=y,y error=error] {
x  y            error
5	2.292       0.083
10	2.773       0.121
20	3.703       0.192
30	4.563       0.374
40	5.315       0.428
50	5.874       0.608
};
\addlegendentry{OpenQASM-total};

\end{semilogyaxis}
\end{tikzpicture} 

\caption{Processing time of Trotter circuits for Heisenberg Hamiltonian models (eq.~\ref{eq:Heisenberg_model}) with a variable number of qubits. Transpiler optimization level 3 and full peephole optimization are applied for Qiskit and t$|$ket$\rangle$, respectively. For OpenQASM compilation, all MLIR-level optimization passes (see~\ref{sec:opt_passes}) and LLVM \texttt{-O3} optimization are applied. For Q\#, QIR generation mode is used to generate LLVM IR, which is then optimized (\texttt{-O3}) by LLVM's \texttt{clang}. Link time to \texttt{qcor}'s QIR runtime is included for both OpenQASM and Q\# compilation time. Each data point is the average of 10 runs with standard deviation also plotted. OpenQASM total time includes both compilation time and resources estimation runtime (counting all executed gates).}
\label{fig:benchmark_chart}
\end{figure}
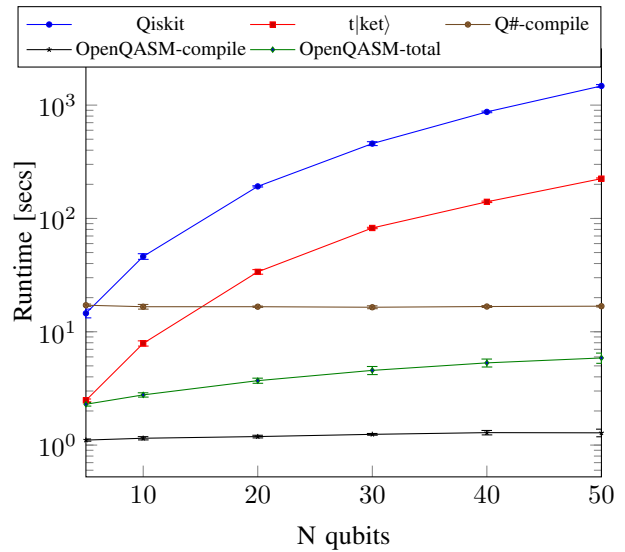

In Figure~\ref{fig:benchmark_chart}, we plot the compile data for Trotter circuits simulating the generic Heisenberg Hamiltonian model of the form 
\begin{equation}
H = -h\sum_i X_i - J_z \sum_i Z_{i}Z_{i+1}.
\label{eq:Heisenberg_model}
\end{equation}
The circuit is constructed by `for' loops over a fixed number of Trotter steps (steps = 100, step size ($dt$) = 0.01) with a variable number of qubits from 5 to 50. In other words, we construct the circuit representing the unitary 
\begin{equation}
U = \prod_{k=1}^{100} \Bigl( \prod_i \exp(-h dt X_i) \prod_i \exp(- J_z dt Z_{i}Z_{i+1})\Bigr),
\label{eq:heisenberg_u}
\end{equation}
where each Pauli exponential term is converted to an equivalent gate-based sub-circuit. These sub-circuits are repeated at each time step to simulate the Trotterized evolution of the Heisenberg Hamiltonian. 

Quantum programming languages, such as Q\# and OpenQASM, preserve the loop construct in their IR representation, resulting in almost constant compilation time. We note that the number of qubits is a compile-time constant for both the Q\# and OpenQASM cases. The compiler may choose to unroll these loops. 

The compilation time of OpenQASM includes: (1) frontend parsing (ANTLR), (2) MLIR generation and optimization, (3) lowering to LLVM IR, (4) LLVM optimization, and (5) object code generation and linking. For Q\#, we leverage the Microsoft Quantum Development Kit (QDK) to perform QIR LLVM generation, i.e., equivalent to steps (1)-(3) in our OpenQASM workflow. The QDK-generated LLVM IR is then optimized and linked with the \texttt{qcor} runtime similar to steps 4 and 5 using the standard LLVM toolchain.

The results in Figure~\ref{fig:benchmark_chart} highlight the need for statically-compiled quantum programming languages in order to describe large-scale programs. Imperative gate-by-gate construction of quantum circuits using scripting languages, despite its flexibility and ease of use, does come with a significant performance overhead. Importantly, our MLIR-based compiler for OpenQASM demonstrates improved performance compared to other compilers, such as Q\#. It is worth noting that the Q\# language is much more feature-rich than OpenQASM, therefore requiring a more elaborated frontend and build system. In particular, initial Q\# to QIR generation accounts for the majority (80-90\%) of the total Q\# compilation time in Figure~\ref{fig:benchmark_chart}. As of this writing, there are no other publicly available OpenQASM compilers that we can compare our implementation with. 

\subsection{Extensions for Optimal Code Generation}
\begin{figure}[t]
\begin{minted}[frame=single,framesep=2pt, fontsize=\footnotesize, breaklines, escapeinside=||, linenos, numbersep=1mm]{llvm}
const nb_qubits = 5;
def heisenberg_U() qubit[nb_qubits]:r {

  // Extension-provided C-like data types
  int nb_steps = 100;
  double step_size = .01;
  double Jz = 1.0;
  double h = 1.0;

  // -h*sigma_x layers
  for step in [0:nb_steps] {
    // -h*sigma_x layers
    rx(-h * step_size) r;

    // -Jz*sigma_z*sigma_z layers
    for i in [0:nb_qubits-1] {
      compute {
        cx r[i], r[i+1];
      } action {
        rz(-Jz * step_size) r[i + 1];
      }
      // Could be written manually like this
      // No optimizations picked up
      // cx r[i], r[i+1];
      // rz(-Jz * step_size) r[i + 1];
      // cx r[i], r[i+1];
    } 
  }
}

// Allocate the qubits
qubit r[nb_qubits], c;

// Perform ctrl-U
ctrl @ heisenberg_U c, r;
\end{minted}
\caption{OpenQASM code defining a subroutine describing the unitary operation in Eq. \ref{eq:heisenberg_u}. The use of \texttt{compute-action} from our grammar extension enables the compiler implementation to optimally synthesis controlled versions of this subroutine. This code snippet also demonstrates the C-like extensions for primitive data types and \texttt{for} loops.}
\label{fig:demo_compute_action}
\end{figure}
Here we demonstrate the utility of our proposed extensions to the OpenQASM grammar specification with regards to optimal quantum code generation. Specifically, we show how the \texttt{compute}-\texttt{action} syntax enables the compiler implementation to generate optimal quantum instruction sequences in the presence of a \texttt{ctrl} gate modifier. As stated in Sec. \ref{sec:extension}, controlled operations on the $W = U V U^\dagger$ pattern only require controls applied to the operations in $V$. Via programmer intent --- i.e. leveraging the custom \texttt{compute \{...\} action \{...\}} syntax --- the compiler can optimally synthesize quantum instructions adherent to this pattern. 

Take the Heisenberg Hamiltonian and corresponding time evolution operator $U$ in Eqs. \ref{eq:Heisenberg_model} and \ref{eq:heisenberg_u}. In the context of the quantum phase estimation algorithm, if we seek a corresponding eigenvalue of $U$ with respect to some eigenstate $|\psi\rangle$, we will require the application of a series of controlled versions of $U$. 

\begin{figure}[t!]
\centering
\begin{tikzpicture}[every mark/.append style={mark size=1pt}]
\begin{semilogyaxis}[
      legend columns=2,
      legend style={at={(0.85,1.2)},anchor=north east},
      xmin = 5, xmax = 50,
      xlabel = {N  Qubits}, 
      ylabel = {N Controlled Operations (CX, CRZ)}, 
      y label style={at={(axis description cs:0.1, 0.55)},anchor=south}]


\addplot+ [error bars/.cd, 
    y fixed,
    y dir=both, 
    y explicit]
table[x=x, y=y] {
x  y  
6 7700
8 10700
10 13700
12 16700
14 19700
16 22700
18 25700
20 28700
22 31700
24 34700
26 37700
28 40700
30 43700
32 46700
34 49700
36 52700
38 55700
40 58700
42 61700
44 64700
46 67700
48 70700
50 73700
};
\addlegendentry{Manual};

\addplot+ [error bars/.cd, 
    y fixed,
    y dir=both, 
    y explicit]
table[x=x, y=y] {
x  y  
6 1700
8 2300
10 2900
12 3500
14 4100
16 4700
18 5300
20 5900
22 6500
24 7100
26 7700
28 8300
30 8900
32 9500
34 10100
36 10700
38 11300
40 11900
42 12500
44 13100
46 13700
48 14300
50 14900
};
\addlegendentry{Compute / Action};
\end{semilogyaxis}
\end{tikzpicture} 

\caption{The number of controlled operations (CX and CRZ) present in the compiled representation of Figure \ref{fig:demo_compute_action} for a given number of qubits. The \emph{Manual} plot represents a program that sequentially lists the instruction for the pattern $W = U V U^\dagger$, while the other plot represents the case whereby a programmer expresses this pattern via the \texttt{compute-action} syntax from our grammar extension.}
\label{fig:benchmark_compute_action_data}
\end{figure}
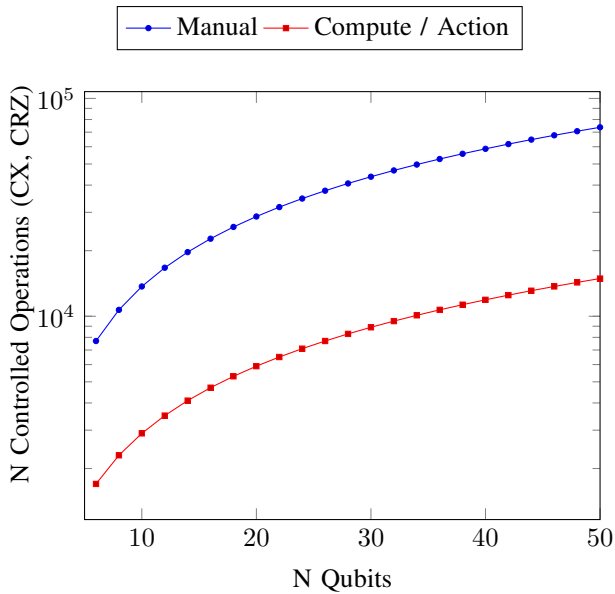
Figure \ref{fig:demo_compute_action} shows an OpenQASM subroutine describing the trotter evolution in Eq. \ref{eq:heisenberg_u}. The second nested \texttt{for} loop (lines 18-29) could be written manually as the sequence $W = CX \otimes RZ(\Theta) \otimes CX^\dagger$, but by replacing it with a \texttt{compute-action} block, we give the compiler an opportunity for optimal instruction synthesis under application of a \texttt{ctrl} modifier. Specifically, we have implemented an ANTLR visitor handler that processes the \texttt{compute-action} source and adds the \texttt{compute} instructions, the \texttt{action} instructions, and the adjoint or reverse of the \texttt{compute} instructions to the MLIR tree. Moreover, for instructions that are not in the \texttt{action} block, the compiler marks the added instructions with a flag to indicate that they are part of the \texttt{compute} or \texttt{uncompute} block. At runtime, this information is used to optimally synthesize controlled versions of this block of code. 

We benchmark the usage of \texttt{compute-action} vs manual programming of $W$ and present the results in Fig \ref{fig:benchmark_compute_action_data}. The results show the number of controlled operations (CRZ, CNOT) present in the compiled quantum program for the manual (commented lines 22-26) and compute-action (lines 17-21) cases. One can clearly see that via programmer intent, the compiler can optimally synthesize instruction sequences and improve on the resource utility of the compiled program. With this simple programming extension, programmers can pick up an order of magnitude in gate count reductions.

\section{Conclusion}
We have presented an optimizing ahead-of-time OpenQASM compiler built on the MLIR framework. Our approach lowers OpenQASM codes to the LLVM IR in a manner that is adherent to the QIR specification. We provide quantum circuit optimization passes at the MLIR level and leverage existing classical optimization passes present in both MLIR and LLVM. Our work extends the OpenQASM grammar with support for C-like constructs and the compute-action-uncompute pattern for efficient programming and compile-time optimizations. Targeting the QIR enables one to swap runtime library implementations enabling a write once run anywhere characteristic. We have provided a runtime library implementation of the QIR specification that is backed by the XACC quantum programming framework, thereby enabling OpenQASM compilation that targets quantum computers from IBM, Rigetti, Honeywell, IonQ, as well as simulators that scale from laptops to large-scale heterogeneous high performance computers, like Summit. Moving forward, our approach opens up the possibility of true language integration at the LLVM IR level. We envision a number of language approaches that map to the LLVM IR adherent to the QIR specification, and via simple runtime linking, enabling the integration of quantum language A with code from quantum language B. As of this writing, the integration of Q\#, \texttt{qcor} C\texttt{++}, and OpenQASM is now possible. We also envision this work as an alternative mechanism for embedded C\texttt{++} quantum kernels in \texttt{qcor}, departing from the existing Clang Syntax Handler source preprocessing. Future work will investigate true quantum language integration and compile-time embedding of MLIR-to-LLVM processing in the \texttt{qcor} C\texttt{++} language extension.

\section*{Acknowledgment}
This material is based upon work supported by the U.S. Department of Energy, Office of Science, National Quantum Information Science Research Centers. ORNL is managed by UT-Battelle, LLC, for the US Department of Energy under contract no. DE-AC05-00OR22725.

\bibliographystyle{plain}
\bibliography{main}

\appendices
\section{Benchmark Experimental Setup}
\label{appendix:benchmark_code}
Here, we list all the source codes used for the Trotter circuit benchmarking (Figure~\ref{fig:benchmark_chart}). The number of qubits in the Q\# and OpenQASM source codes represent a particular data point. We modify the number of qubits and recompile the source code for the benchmark.
\subsection{Qiskit script}
\begin{minted}[frame=single,framesep=2pt, fontsize=\footnotesize, breaklines, escapeinside=||, linenos, numbersep=1mm]{python}
from qiskit.aqua.operators import (X, Z, I, EvolvedOp, PauliTrotterEvolution)
from qiskit import QuantumRegister, QuantumCircuit
from qiskit.compiler import transpile
from statistics import mean, stdev
import time

def X_op(idxs, n_qubits):
  op = None
  if 0 in idxs:
    op = X
  else:
    op = I
  for i in range(1, n_qubits):
    if (i in idxs):
      op ^= X
    else:
      op ^= I
  return op

def Z_op(idxs, n_qubits):
  op = None
  if 0 in idxs:
    op = Z
  else:
    op = I
  for i in range(1, n_qubits):
    if (i in idxs):
      op ^= Z
    else:
      op ^= I
  return op

def heisenberg_ham(n_qubits):
  Jz = 1.0
  h = 1.0
  H = -h * X_op([0], n_qubits)
  for i in range(1, n_qubits):
    H = H - h * X_op([i], n_qubits)
  for i in range(n_qubits - 1):
    H = H - Jz * (Z_op([i, i + 1], n_qubits))
  return H

n_qubits = [10, 20, 50, 100]
nbSteps = 100
n_runs = 10

def trotter_circ(q, exp_args, n_steps):
  qc = QuantumCircuit(q)
  for i in range(n_steps):
    for sub_op in exp_args:
      qc += PauliTrotterEvolution()
            .convert(EvolvedOp(sub_op))
            .to_circuit()
  return qc

for nbQubits in n_qubits:
  data = []
  for run_id in range(n_runs):  
    ham_op = heisenberg_ham(nbQubits)
    q = QuantumRegister(nbQubits, 'q')
    start = time.time()
    comp = trotter_circ(q, ham_op.oplist, nbSteps)
    comp = transpile(comp, optimization_level=3)
    end = time.time()
    data.append(end - start)
  print('n_qubits =', nbQubits, '; Elapsed time =', mean(data), '+/-', stdev(data),  '[secs]')
\end{minted}
\subsection{t$|$ket$\rangle$ script}
\begin{minted}[frame=single,framesep=2pt, fontsize=\footnotesize, breaklines, escapeinside=||, linenos, numbersep=1mm]{python}
import time
from pytket.circuit import Circuit, PauliExpBox
from pytket.pauli import Pauli
from pytket.extensions.qiskit import AerStateBackend
from pytket.passes import FullPeepholeOptimise
from statistics import mean, stdev
nb_steps = 100
step_size = 0.01
n_qubits = [5, 10, 20, 30, 40, 50]
n_runs = 10
for nb_qubits in n_qubits:
  data = []
  for run_id in range(n_runs):  
    # Start timer
    start = time.time()
    circ = Circuit(nb_qubits)
    h = 1.0
    Jz = 1.0
    for i in range(nb_steps):
        # Using Heisenberg Hamiltonian:
        for q in range(nb_qubits):
            circ.add_pauliexpbox(PauliExpBox( [Pauli.X], -h * step_size), [q])
        for q in range(nb_qubits - 1):
            circ.add_pauliexpbox(PauliExpBox( [Pauli.Z, Pauli.Z], -Jz * step_size), [q, q + 1])

    # Compile to gates
    backend = AerStateBackend()
    circ = backend.get_compiled_circuit(circ)

    # Apply optimization
    FullPeepholeOptimise().apply(circ)

    end = time.time()
    data.append(end - start)
  
  print('n_qubits =', nb_qubits, '; Elapsed time =', mean(data), '+/-', stdev(data),  '[secs]')
\end{minted}
\subsection{Q\# source code}
\begin{minted}[frame=single,framesep=2pt, fontsize=\footnotesize, breaklines, escapeinside=||, linenos, numbersep=1mm]{csharp}
namespace Benchmark.Heisenberg {
    open Microsoft.Quantum.Intrinsic;
    open Microsoft.Quantum.Canon;
    open Microsoft.Quantum.Math;
    open Microsoft.Quantum.Convert;
    open Microsoft.Quantum.Arrays;
    open Microsoft.Quantum.Measurement;
    // In this example, we will show how to simulate the time evolution of
    // an Heisenberg model: 
    operation HeisenbergTrotterEvolve(nSites : Int, simulationTime : Double, trotterStepSize : Double) : Unit {
        // We pick arbitrary values for the X and J couplings
        let hXCoupling = 1.0;
        let jCoupling = 1.0;

        // This determines the number of Trotter steps
        let steps = Ceiling(simulationTime / trotterStepSize);

        // This resizes the Trotter step so that time evolution over the
        // duration is accomplished.
        let trotterStepSizeResized = simulationTime / IntAsDouble(steps);

        // Let us initialize nSites clean qubits. These are all in the |0>
        // state.
        use qubits = Qubit[nSites];
        // We then evolve for some time
        for idxStep in 0 .. steps - 1 {
            for i in 0 .. nSites - 1 {
                Exp([PauliX], (-1.0 * hXCoupling) * trotterStepSizeResized, [qubits[i]]);
            }
            for i in 0 .. nSites - 2 {
                Exp([PauliZ, PauliZ], (-1.0 * jCoupling) * trotterStepSizeResized, qubits[i .. (i + 1)]);
            }
            
        }
    }

    // Entry point: we allow the Q# program to have full information (compile-time) about the number of qubits, steps, etc.
    @EntryPoint()
    operation CircuitGen() : Unit {
        HeisenbergTrotterEvolve(50, 1.0, 0.01);
    }
}
\end{minted}
\subsection{OpenQASM source code}
\begin{minted}[frame=single,framesep=2pt, fontsize=\footnotesize, breaklines, escapeinside=||, linenos, numbersep=1mm]{asm}
OPENQASM 3;

const nb_steps = 100;
const nb_qubits = 50;
const step_size = 0.01;
const Jz = 1.0;
const h = 1.0;

qubit r[nb_qubits];
for step in [0:nb_steps] {
  // -h*sigma_x layers
  for i in [0:nb_qubits] {
    rx(-h * step_size) r[i];
  }

  // -Jz*sigma_z*sigma_z layers
  for i in [0:nb_qubits - 1] {
    cx r[i], r[i+1];
    rz(-Jz * step_size) r[i + 1];
    cx r[i], r[i+1];
  }
}
\end{minted}
\end{document}